\documentclass[9pt, twocolumn]{extarticle}

\usepackage{hyperref}
\usepackage{breakurl} 
\usepackage{amsmath}
\usepackage{amsfonts}
\usepackage{amsthm}
\usepackage{amssymb}
\usepackage[a4paper, twocolumn, left=2cm, right=2cm,columnsep=1cm,textheight=22cm]{geometry}
\usepackage{xcolor}
\usepackage{lstcoq}
\usepackage{todonotes}
\usepackage{paralist}

\renewenvironment{itemize}[1]{\begin{compactitem}#1}{\end{compactitem}}

\begin{document}
\title{Category Theory in Coq 8.5}
\author{Amin Timany \hspace{2em} Bart Jacobs \\[1em]
iMinds-DistriNet -- KU Leuven \\[0.2em]
firstname.lastname@cs.kuleuven.be}
\date{}

\maketitle

We report on our experience implementing category theory in Coq 8.5
\footnote{The last tested version is Coq 8.5-beta1 but we intend to keep compatibility at least up to the official release of Coq 8.5}.
The repository of this development can be found at \url{https://bitbucket.org/amintimany/categories/}.
This implementation most notably makes use of features \emph{primitive projections} for records and \emph{universe polymorphism} that are new to Coq 8.5.
The latter provides relative smallness and largeness in the development.
This will be elaborated below.
The former allows for specification of well-behaved dualities in the category theoretical sense.
That is, we get definitional equalities such as\footnote{This was achieved similarly to the implementation of category theory on top of Coq/HoTT \cite{DBLP:conf/itp/GrossCS14}}:
\[
(C^{op})^{op} = C \hspace{2em} (F^{op})^{op} = F \hspace{2em} (N^{op})^{op} = N
\]
\[
(F \circ F')^{op} = F^{op} \circ F'^{op}
\]
Where $C$ is a category, $F$ and $F'$ are functors and $N$ is a natural transformation.

In this development the category \texttt{Type\_Cat} plays the role of category of sets \textbf{Set}. This is the category of types in Coq as objects and functions among them as arrows. In the sequel, we will simply use \textbf{Set} to denote this category.

The following is the list of the most important notions and features of this development.
\begin{itemize}[\tiny$\blacksquare$]
\item basic constructions:
	\begin{itemize}
	 \item terminal/initial object
	 \item products/sums
	 \item equalizers/coequalizers
	 \item pullbacks/pushouts
	 \item exponentials
	 \item $+ \dashv \Delta \dashv \times$ and $(- \times a) \dashv a^-$
	\end{itemize}
\item external constructions:
	\begin{itemize}
	 \item comma categories
	 \item product category
	\end{itemize}
\item for \textbf{Cat}: (\Coqe|Obj := Category, Hom := Functor|)
	\begin{itemize}
	 \item cartesian closure
	 \item initial object
	\end{itemize}
\item for \textbf{Set}: (\Coqe|Obj := Type, Hom := fun A B => A -> B|)
	\begin{itemize}
	 \item initial object
	 \item sums
	 \item equalizers
	 \item coequalizers\textsuperscript{$\dagger$}
	 \item pullbacks
	 \item cartesian closure
	 \item local cartesian closure\textsuperscript{$\dagger$}
	 \item completeness
	 \item co-completeness\textsuperscript{$\dagger$}
	 \item sub-object classifier (\Coqe|Prop : Type|)\textsuperscript{$\dagger$}
	 \item topos\textsuperscript{$\dagger$}
	\end{itemize}
\item the Yoneda lemma
\item adjunction
	\begin{itemize}
	\item hom-functor adjunction, unit-counit adjunction, universal morphism adjunction and their conversions
	\item duality : $F \dashv G \Rightarrow G^{op} \dashv F^{op}$
	\item uniqueness up to natural isomorphism
	\end{itemize}
\item kan extensions
	\begin{itemize}
	\item global definition
	\item local definition with both hom-functor and cones (along a functor)
	\item uniqueness
	\item preservation by adjoint functors
	\item pointwise kan extensions (preserved by representable functors)
	\end{itemize}
\item (co)limits
	\begin{itemize}
	\item as (left)right local kan extensions along the unique functor to the terminal category
	\item (sum)product-(co)equalizer (co)limits
	\item pointwise (as kan extensions)
	\end{itemize}
\item $T-(co)algebras$ (for an endofunctor $T$)
\item[] \textsuperscript{$\dagger$}uses the axioms of propositional extensionality and constructive indefinite description (axiom of choice).
\end{itemize}

In addition, we have used the axiom of functional extensionality and proof-irrelevance.
The axiom of proof-irrelevance is mostly used in proofs of equality of arrows, e.g., to prove two functors are equal, one just needs to prove the object- and arrow-maps are equal.

Even though (co)limits are defined in general, we have defined most important and useful (co)limits separately: terminal object, products, equalizers and pullbacks.
The duals of these notions, i.e., initial object, sums, coequalizers and pushouts, respectively, are simpliy defined as their counterparts in the opposite category.
Similarly, only the right local kan extensions (in both versions) are defined directly and local left can extensions are simply assumed as local right kan extensions with opposite functors.

Although dualities behave nicely (in the aforementioned sense), working with dual definitions is not always as smooth.
This is especially evident in rewriting equalities.
In some cases one has to add the equality to the proof context (usually applied to the arguments that are difficult to match) and perform a simplification on them before they can be used with the \Coqe|rewrite| tactic.
In some rare extreme cases, simplifications with tactics like \Coqe|cbn| and \Coqe|simpl| are not enough and one has to change the goal in such a way that those lemmas can be used with, e.g., the \Coqe|apply| tactic, instead of \Coqe|rewrite|.

\subsubsection*{Universe levels, smallness and largeness}
In this implementation, we use universe levels as the underlying notion of smallness/largeness.
In other words, each category has universe levels that indicate its relative smallness/largeness.
In practice, the type of categories has two universe level parameters, \Coqe|Category@{i,j} : Type|\Coqe|@{max(i+1, j+1)}|, where \texttt{i} is the level of the type of objects and \texttt{j} is the level of the type of arrows.
This relative notion of smallness/largeness works so well, in fact, that we can prove the following theorem in our implementation:
\begin{Coq}
Theorem Complete_Preorder (C : Category) (CC : Complete C) :
   forall x y : (Obj C), Hom x y' $\simeq$  ((Arrow C) -> Hom x y)
\end{Coq}
where \texttt{y'} is the limit of the constant functor from the discrete category \texttt{(Arrow C)} that maps every object to \texttt{y} and \texttt{(Arrow C)} is the type of all homomorphisms of category \texttt{C}.
This though, would result in a contradiction as soon as we have two objects \texttt{c} and \texttt{d} in \texttt{C} for which \texttt{Hom c d} has more than one element.
That is, we have effectively shown that \emph{any} complete category is a preorder category, i.e., between any two objects there is at most one arrow.
This is indeed absurd as the category \textbf{Set} is complete and there are types in Coq that have more than one function between them! 
However, this theorem holds for small categories.
That is, any \emph{small} and complete category is a preorder category\footnote{This theorem and its proof are taken from \cite{awodey2010category}.}.
Expectedly, the restrictions on the universe levels of this theorem do indeed confirm this fact.
That is, this theorem is in fact only applicable to categories in which the level of the type of objects is less than or equal to the level of the type of arrows.
Hence, Coq will refuse to apply this theorem to the category \textbf{Set} with a universe inconsistency error as for it the level of the type of arrows is strictly less than the level of the type of objects.

On the other hand, the universe polymorphism of Coq treats inductive types by considering copies of them at different levels. See \cite{DBLP:conf/itp/SozeauT14}.
That implies that if we have \Coqe|C : Category@{i, j}| and we additionally have that \Coqe|C : Category@{i', j'}|, Coq enforces \texttt{i} $=$ \texttt{i'} and \texttt{j} $=$ \texttt{j'}.
In this setting, the category of (relatively small) categories \textbf{Cat}, which in the implementation has type
\begin{Coq}
Cat@{i, j, k, l} : Category@{i, j}
\end{Coq}
is \emph{not} the category of all smaller categories.
Rather it is the category of categories that are at level \texttt{k} and \texttt{l} and not any lower level.

Apart from the fact that \textbf{Cat} defined this way is not the category of all relatively small categories, these restrictions on universe levels impose practical restrictions as well.
For instance, looking at the fact that \Coqe|Cat@{i, j, k, l}| has exponentials (functor categories), we can see the restriction that \texttt{j} $=$ \texttt{k} $=$ \texttt{l}.
That is only those copies have exponentials for which this restriction holds.
Looking back at the category of types, \textbf{Set}, we had the restriction that the level of the type of arrows is strictly less than that of objects.
This means, there is no version of \textbf{Cat} that both has exponentials and a version of \textbf{Set} in its objects.

This means that we can't simply assume that \textbf{Cat} has exponentials and get the exponential transpose of the hom-functor to be the Yoneda embedding\footnote{However, the definition of currying defined \emph{independently of the notion of exponentials} for functors can be used to this end and it is precisely how we have defined the Yoneda embedding.}.

Moreover, we can use the Yoneda lemma to show that in any cartesian closed category, for any objects $a, b$ and $c$:
\[
{(a^b)}^c \simeq a^{b \times c}
\]
Yet, this theorem can't be applied to \textbf{Cat}, even though it holds for \textbf{Cat}.

On the other hand, if we show that \textbf{Set} has the type \Coqe|unit| as its terminal object, we, strangely, get the restriction that the level of the type of arrows of \textbf{Set} is universe \Coqe|Set| but, expectedly, not for objects.
A similar problem happens with showing that the category whose object type and arrow type are \Coqe|unit| is the terminal object of \textbf{Cat}.
It is not clear to the authors wether this is intensional or the result of a bug.
In any case, we have elected to go around this problem by postulating existence of a universe polymorphic type that has a single inhabitant:
\begin{Coq}
Parameter UNIT : Type.
Parameter TT : UNIT.
Axiom UNIT_SINGLETON : forall x y : UNIT, x = y.
\end{Coq}

\bibliography{ref}{}
\nocite{*}
\bibliographystyle{plain}

\end{document}